# Telecom-wavelength Single-photon Emitters in Multi-layer InSe

Huan Zhao[1,2,*], Saban Hus[1], Jinli Chen[3], Xiaodong Yan[3], Ben Lawrie[1,4], Stephen Jesse[1], An-Ping Li[1], Liangbo Liang[1,*], Han Htoon[2,*]

[1]Center for Nanophase Materials Sciences, Oak Ridge National Laboratory

[2]Center for Integrated Nanotechnologies, Los Alamos National Laboratory

[3]Department of Materials Science and Engineering, University of Arizona

[4]Materials Science and Technology Division, Oak Ridge National Laboratory

* Corresponding authors, Email: zhaoh1@ornl.gov, liangl1@ornl.gov, htoon@lanl.gov

**Abstract**

The development of robust and efficient single photon emitters (SPEs) at telecom wavelengths is critical for advancements in quantum information science. Two-dimensional (2D) materials have recently emerged as promising sources for SPEs, owing to their high photon extraction efficiency, facile coupling to external fields, and seamless integration into photonic circuits. In this study, we demonstrate the creation of SPEs emitting in the 1000 to 1550 nm near-infrared range by coupling 2D indium selenide (InSe) with strain-inducing nanopillar arrays. The emission wavelength exhibits a strong dependence on the number of layers. Hanbury Brown and Twiss experiments conducted at 10 K reveal clear photon antibunching, confirming the single-photon nature of the emissions. Density-functional-theory calculations and scanning-tunneling-microscopy analyses provide insights into the electronic structures and defect states, elucidating the origins of the SPEs. Our findings highlight the potential of multilayer 2D metal monochalcogenides for creating SPEs across a broad spectral range, paving the way for their integration into quantum communication technologies.

**Main Text**

Solid-state single photon emitters (SPEs) are essential for advancements in quantum information science. These quantum light sources play a crucial role in encoding and transmitting quantum information via photons, serving as the foundation for technologies poised to revolutionize computing, secure communication, and sensing. [1,2] SPEs that operate at near-infrared (NIR) wavelengths, particularly those within the telecom band, hold significant technological importance. [3-5] The alignment of these SPEs with established fiber optic technologies - characterized by minimal photon loss at the telecom wavelengths[6] - facilitates the seamless integration of quantum and classical communication infrastructures. Presently, only a limited number of materials are known to host telecom-band SPEs (Table S1). [7-13] Expanding the range of materials capable of telecom-band quantum emission will create new opportunities in quantum information science (QIS) research.

Recent advancements have demonstrated the potential of solid-state SPEs in two-dimensional (2D) materials (Supplementary Note 1). These 2D material-based SPEs includes hexagonal boron nitride (hBN), [14,15] GaSe, [16-18] transition metal dichalcogenides (TMDCs) such as $WSe_2$, [19-22] $MoTe_2$, [8] and TMDC heterostructures. [23,24] The TMDCs exhibit a direct bandgap across the K valleys in monolayer forms, which shifts to an indirect bandgap as the material thickness increases. The SPEs in these materials typically originate from bound-state excitons in the K-band, leveraging the direct inter-band transitions to produce bright emissions in monolayers. [25] In contrast, the emissions become increasingly dim in multilayer and bulk forms due to the less efficient indirect bandgap transitions. Therefore, the ability to tune SPE wavelengths by varying the number of layers is constrained by the weak emission from thick samples. Moreover, the use of monolayers presents practical challenges, as they are more difficult to fabricate and more susceptible to environmental degradation than thicker samples. In contrast, group III–VI metal monochalcogenides (MMCs), such as InSe and GaSe, exhibit a direct bandgap for multi-layer and bulk samples. [26, 27] The bandgap, which is strongly layer-dependent, converges at Γ band edges. [28, 29] This characteristic enables effective tuning of the emission wavelength through layer manipulation, offering a versatile approach to developing bright SPEs that are robust, tunable, and compatible with multilayer structures, thereby expanding the functional scope of photon emitter applications.

In this study, we demonstrate the site-controlled creation of telecom-band SPEs using layered InSe, a 2D MMC characterized by Γ band emission. By applying strain and manipulating the number of layers, we achieved bright, localized emissions spanning 1000 to 1550 nm. These localized emissions are primarily attributed to the strain-induced modification of the band alignment as revealed by first-principles density-functional-theory (DFT) calculations. Within the strain profile, in-gap energy levels arising from selenium vacancies could hybridize with the strain-modified band edges to serve as recombination centers for the localized excitons.

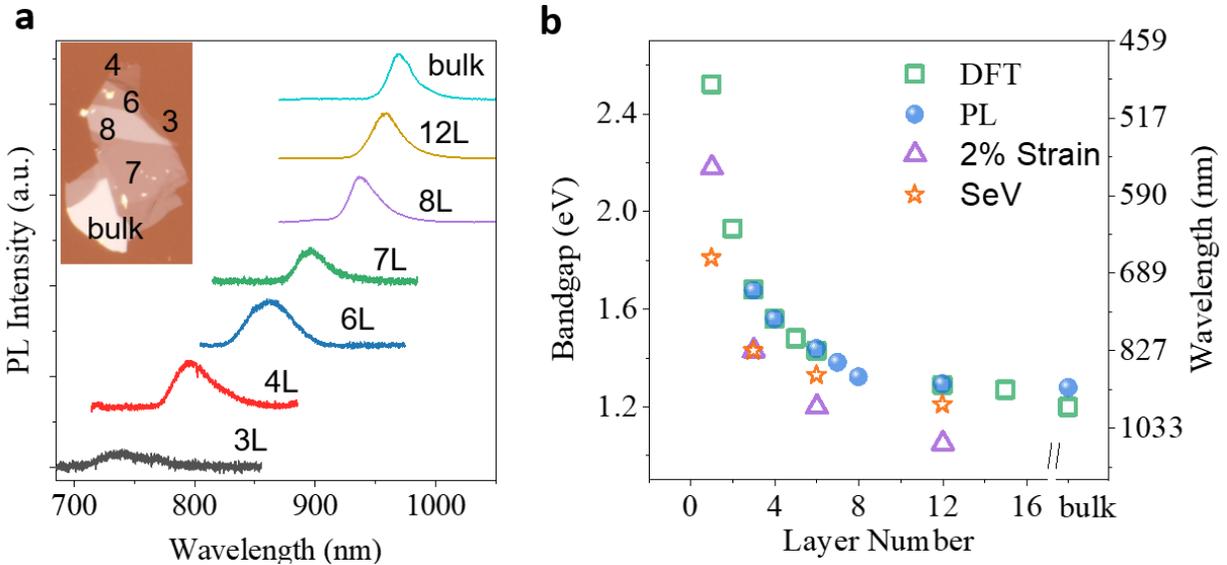

*Figure 1: Layer-dependent bandgap in multilayer InSe*. **a**, Layer-dependent PL at 10 K temperature. The flakes with 3-6 layers were measured using a silicon CCD, while all other samples were measured using an InGaAs camera. **b**, Comparison of layer-dependent bandgap extracted from PL peaks (blue spheres) and from DFT calculations (green boxes). The calculated layer-dependent bandgap with 2% biaxial strain (purple triangles) and with Se vacancy defect (orange stars) are also displayed. All PL measurements were performed at 10 K temperature.

The layer-dependent behavior of as-exfoliated InSe layers is revealed in Figure 1a by low-temperature photoluminescence (PL) under 800 nm laser excitation; the bandgap progressively reduces from 1.68 eV in trilayers to 1.28 eV in bulk samples. This trend continues even for thicker samples with more than 10 layers. This behavior starkly contrasts with that observed in 2D TMDCs such as $MoS_2$, where the predominant A exciton emission energy only marginally decreases with layer thickness and plateaus within 10 layers. [30, 31] Furthermore, the direct bandgap nature of multilayer InSe leads to enhanced emission brightness as the number of layers increases, attributed to an increase in photocarriers. The DFT-calculated layer-dependent bandgap, shown in Figure 1b, corroborates our experimental observations.

The pronounced layer-dependent emission properties are also observed and even further enhanced in strain-induced-defect-bound localized emitters in multiplayer InSe. Figure 1b displays DFT calculations that predict both strain and defect can narrow the bandgap for monolayer, few-layer, and multi-layer samples (see Figure S5 for calculated band structures). Figure 2a and 2d present optical images of 8.6 nm (10 layer, or 10L) and 34.1 nm (40 layer, or 40L) thick InSe flake positioned on strain-inducing dielectric nanopillar arrays made of polymethyl methacrylate (PMMA). The substrate, comprising gold on silicon, effectively quenches any undesirable off-pillar emissions. Figures 2b to 2d depict the PL intensity images for the 10L sample under various long-pass (LP) optical filters within the collection channel. With an 870 nm LP filter, the PL intensity image captures emissions from both intrinsic 2D excitons and strain-induced localized emissions. Application of a 1064 nm LP filter completely blocks intrinsic emissions, showcasing the strain-enabled localized emissions at redshifted wavelength. Notably, no emission with wavelengths longer than 1200 nm was detected. Conversely, the 40-layer sample, evaluated under identical conditions as the 10-layer sample, displays generally brighter emissions with more significant PL redshift. Strain-induced emissions beyond 1200 nm wavelength are frequently observed, making 40L InSe promising for telecom SPE creation. The PL image and spectra of an approximately 20L InSe flake on a nanopillar array are shown in Supplementary Fig. S2, revealing localized emissions up to 1250 nm.

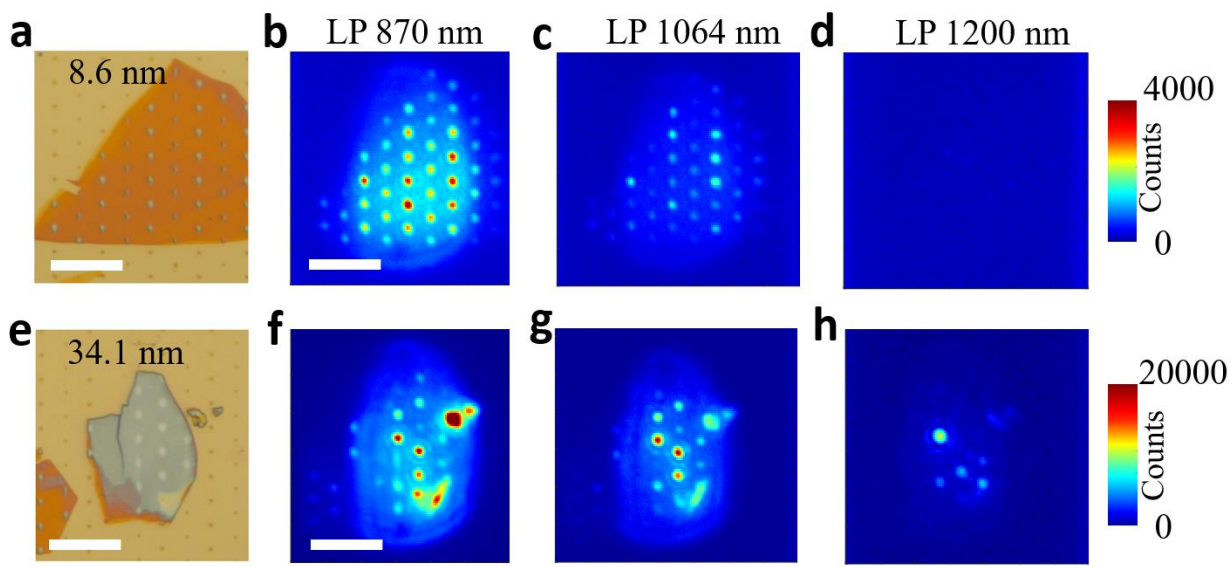

*Figure 2: Strain-induced localized emissions in thick InSe flakes*. (**a**, **e**) Optical images of a 10L InSe flake and a 40L InSe flake, respectively. Both flakes were excited under wide-field illumination using a Gaussian laser beam with a ~30 μm diameter. Various long-pass optical filters were applied to obtain PL intensity images capturing emissions from wavelengths above 870 nm (**b**, **f**), 1064 nm (**c**, **g**), and 1200 nm (**d**, **h**). All samples were measured at 10 K using a liquid-nitrogen-cooled InGaAs camera, sensitive in the 870-1570 nm spectral range. The scale bar is 15 μm for all images.

To evaluate the single-photon-emission characteristics of the InSe localized emitters, we conducted time-tagged, time-correlated single-photon counting alongside Hanbury Brown and Twiss (HBT) experiments. The time-resolved photoluminescence (TRPL) and the second-order correlation function, $g^2(\tau)$, were derived from the photon statistics. Figure 3a illustrates the PL spectrum of a localized emitter in the 10-layer InSe sample measured at 10K temperature, with the corresponding PL decay curve depicted in Figure 3b. This curve exhibits a bi-exponential decay, featuring lifetimes of $\tau_1 = 15.9 \pm 0.1$ ns and $\tau_2 = 119.8 \pm 0.9$ ns. From prior TRPL studies on multi-layer InSe, [28] $\tau_1$ is likely associated with indirect transitions, while $\tau_2$ results from defect-related recombination. The notably prolonged lifetime, relative to the < 1 ns typical of 2D InSe excitons, [32] suggests localization of the exciton within a strain-induced potential trap, effectively mitigating fast non-radiative recombination. Figure 3c shows the second-order correlation function exhibits a prominent dip at zero time delay, with $g^{(2)}(0) = 0.40 \pm 0.01$, evidencing photon antibunching.

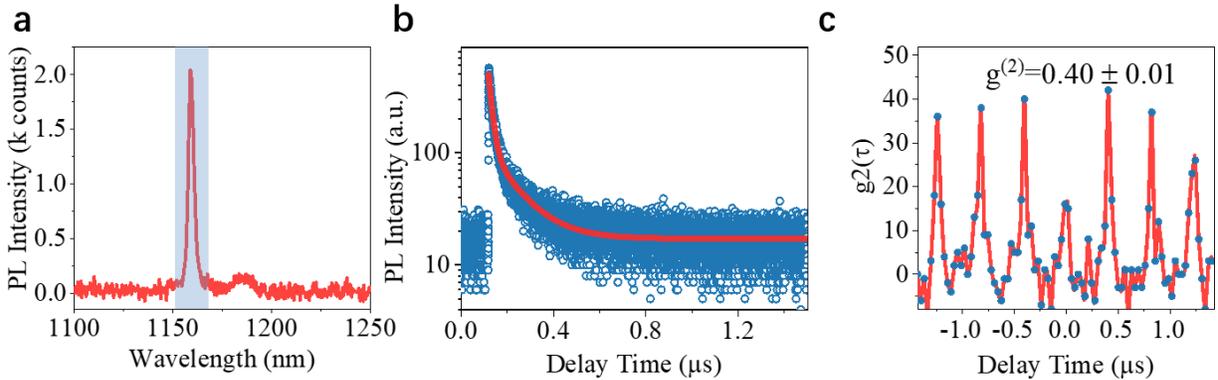

**Figure 3: Photon characterization of a localized emitter in 10L InSe. a**, PL spectrum of a localized InSe emitter. The data of Fig. **b, c** are taken from this SPE with a band-pass filter that allows the shadowed region to be detected. b, The PL decay curve (blue) and a bi-exponential decay fit (red) with $\tau_1$ = 15.9 ± 0.1 ns and $\tau_2$ = 119.8 ± 0.9 ns. c, Second-order correlation measurement under 800 nm pulsed excitation with a 2.4 MHz repetition rate, from which a $g^{(2)}(0)$ = 0.40 ± 0.01 is extracted. All data were obtained at 10 K temperature.

Telecom-compatible quantum emitters that operate around 1.3 μm (O-band, "original band") and 1.55 μm (C-band, "conventional band") are crucial for fiber-based quantum communications. In InSe flakes ranging from 20-40 nm thickness, we frequently observed telecom-band emissions at nanopillar sites. Figure 4a displays the PL spectra of these emitters with emission peaked at 1.30–1.55 μm spectral range. Figure 4b presents the PL spectrum of an O-band telecom SPE, for which the PL dynamics and photon correlation results are presented in Figure 4c–e. TRPL measurements revealed an initial rapid PL decay with a 3.01 ± 0.02 ns lifetime, followed by slower decays of 15.10 ± 0.31 ns and 89.48 ± 1.93 ns (Figure 4c). The time-dependent PL displayed in Figure 4d shows no signs of blinking or photobleaching over an 1800 s duration, confirming the stability of the emitter. HBT experiments under pulsed excitation, as shown in Figure 4e, recorded a $g^{(2)}(0)$ of 0.34 ± 0.01, demonstrating the antibunching nature of these telecom photons. However, not every

InSe localized emitter displayed clear photon antibunching. For instance, the emission peaking at 1450 nm exhibited a $g^{(2)}(0)$ of $0.60 \pm 0.02$ (Supplementary Fig. S3), and the emission at 1550 nm lacked a detectable dip in $g^{(2)}(0)$. These results suggest the presence of multiple emitters or emissive states within the excited spots.

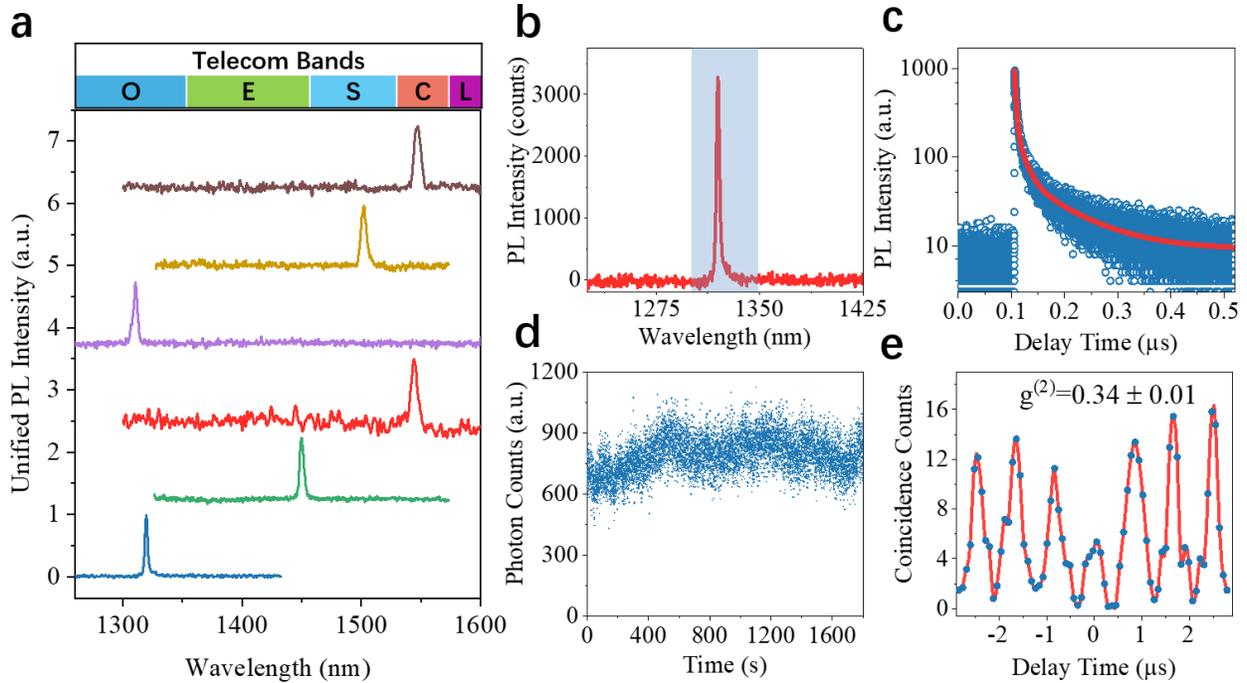

*Figure 4: Photon characterization of a localized emitters in 40L InSe*. **a,** Telecom-wavelength PL emissions from localized emitters. The telecom wavelength bands are classified according to their wavelength ranges and are represented by the letters O, E, S, C, and L. These correspond to the Original band (1260–1360 nm), Extended band (1360–1460 nm), Short-wavelength band (1460–1530 nm), Conventional band (1530–1565 nm), and Long-wavelength band (1565–1625 nm), respectively. **b,** InSe SPE with a telecom O-band emission peak. The data of Fig. c-e are taken from this emitter with a band-pass filter that allows only the shaded region to be detected. **c** PL decay curve (blue) and tri-exponential decay fit (red) reveal the primary $3.01 \pm 0.02$ ns decay followed with two slower decays of $15.10 \pm 0.31$ ns and $89.48 \pm 1.93$ ns. **d,** Time-dependent PL counts showing stable emission over 1800 s. the modulation observed over the long timescale (~100 s) is mainly due to sample drift relative to the laser excitation focal spot. Data were recorded every 100 ms. **e,** Second-order correlation measurement under 800 nm pulsed excitation with a 1.24 MHz repetition rate, from which $g^{(2)}(0) = 0.34 \pm 0.01$ is extracted. All data were obtained at 10 K temperature.

Although the origin of SPEs in 2D semiconductors is still under debate, it is generally believed that both strain and defect states play a crucial role. Strain potential can narrow the bandgap and funnel/localize excitons, while defect levels provide recombination centers for such localized excitons. [33, 34] To study the origin of SPEs in multilayer InSe, we conducted DFT calculations on the electronic structures of intrinsic and 2% strained 12-layer InSe (Figure 5a, 5b). The non-strained 12-layer InSe shows a direct bandgap of approximately 1.29 eV across the Γ band edges. With 2% biaxial tensile strain applied, the conduction band minimum (CBM) downshifts while the valence band maximum (VBM) upshifts, effectively narrowing the bandgap. Figure 5c displays how the CBM and VBM converge with increased strain, culminating in a type-I "staggered" band alignment under localized strain, thereby trapping excitons within the strain center, as illustrated in Figure 5d. The new band edges could hybridize with in-gap defect states to enhance localized emissions.

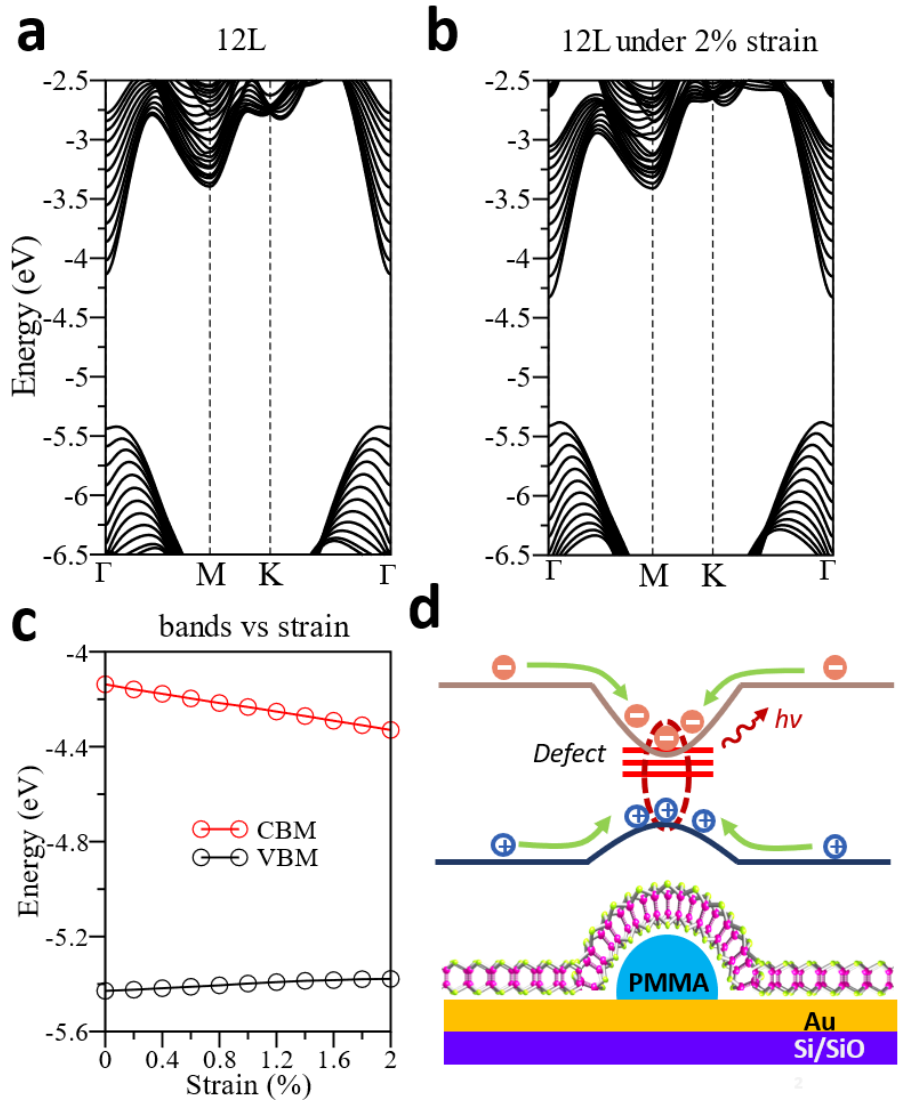

*Figure 5: DFT calculated electronic band structures of 12L InSe*: **a**, no strain and **b**, 2% biaxial tensile strain. All band energies are aligned to the vacuum potential for direct comparison. **c**, The calculated energy positions of the VBM and CBM of 12L InSe as a function of the strain. **d**, Schematic representation of the energy band alignment of InSe on a nanopillar shows a type-I 'staggered' band alignment and potential band mixing between the strain-modulated band edge and selenium defect levels.

To understand how the atomistic defects in InSe affect the SPE creation, we performed scanning tunneling microscope (STM) imaging and scanning tunneling spectroscopy (STS) measurements. High-resolution STM images (Figure 6a) of bulk InSe exfoliated in Ultra High Vacuum (UHV) display a honeycomb-like structure, with three brighter corners corresponding to selenium atoms and three darker corners corresponding to indium atoms. STS conducted in these defect-free regions shows a quasi-particle bandgap of approximately 1.30 ± 0.05 eV (Figure 6b). Considering the ~20 meV exciton binding energy reported in bulk InSe,[35] the optical bandgap of bulk InSe should be ~1.28 ± 0.05 eV, which is consistent with our DFT and PL results. Additionally, STS measurements reveal an n-doped transport character. Large-scale STM images of UHV exfoliated samples indicate the presence of various defect types, with a total defect density of $3\times10^{12}$ cm$^{-2}$ (Figure 6c). Among these defects, the most common are the dark, shallow spots with a diameter of about 4 nm (type 1 defect, D1) and regions with a sharp, bright spot at the center (type 2 defect, D2). The D2 sites can be readily created with the STM tip, suggesting they are vacancies of either indium or selenium atoms. A previous experimental study revealed a preferential formation of selenium vacancies over indium vacancies in InSe,[36] leading us to attribute D2 to native selenium vacancies. Next, we exposed the sample to ambient conditions for one minute to mimic the conditions of samples used in optical measurements. STM images show that the density of D1 sites increased more than tenfold after brief air exposure, while the change in the density of D2 sites was less significant (Figure 6d). Given its spatially broad feature and prevalence in air-exposed samples, we attribute D1 to oxygen substitution at selenium sites, a phenomenon predicted and observed in previous InSe studies.[36-40] Our DFT calculations predict that oxygen substitutions heal the Se vacancies and do not create in-gap states (Supplementary Fig. S4), consistent with previous studies.[36] Conversely, DFT reveals that selenium vacancies can lower the CBM, narrowing the bandgap.[37] This lowered CBM could hybridize with the strain-induced CBM downshift, serving as recombination centers for SPE generation. The presence of Se vacancies in UHV-exfoliated InSe suggests that these defects could emerge spontaneously. However, the random distribution of these vacancies often results in multiple emitters within a single confocal excitation spot, thereby compromising the purity of SPEs. To attain more consistent SPE characteristics, the precise implantation of deep defect states, such as germanium centers,[37] could be employed.

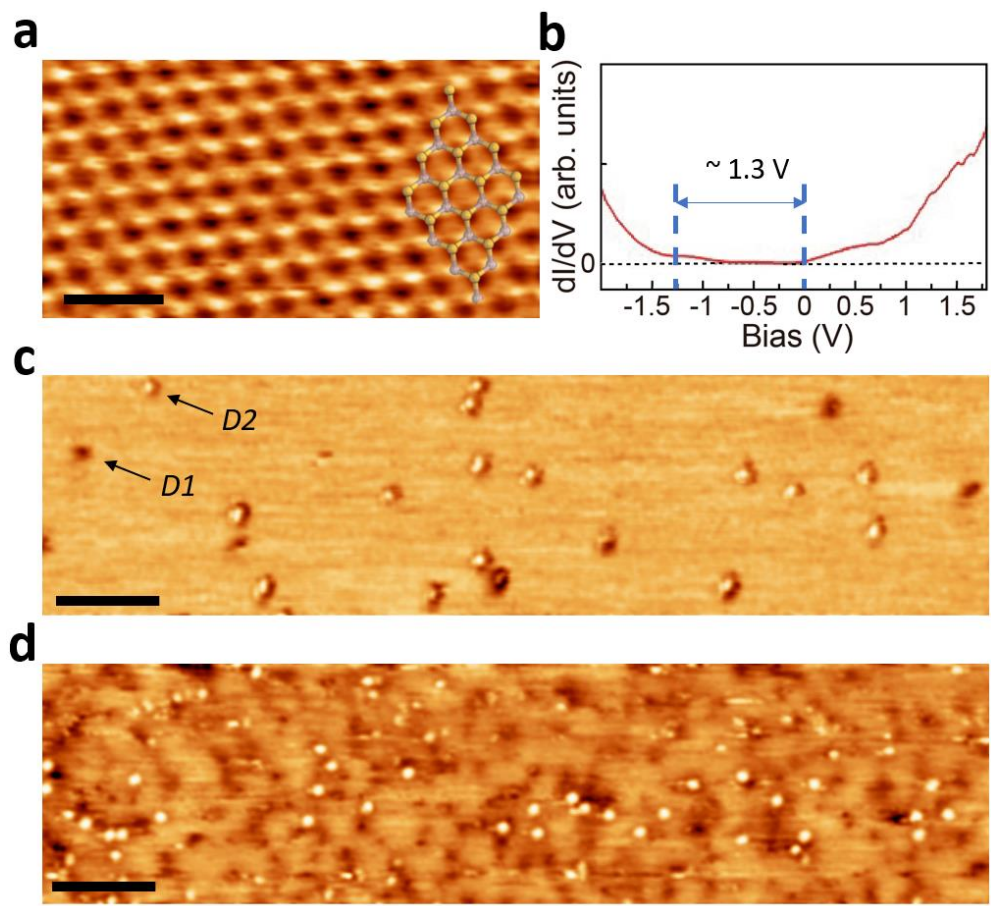

*Figure 6: STM and STS of bulk InSe:* **a**, Atomic-resolution STM image of the InSe surface. In the overlaid ball-and-stick model, yellow spheres represent selenium atoms and grey spheres represent indium atoms. Scale bar: 1 nm. **b**, STS spectra taken from defect-free regions in (a), with dashed lines indicating the band edges. Set points for (a) and (b): 1.5 V, 200 pA. **c**, Large-scale STM image of a UHV-exfoliated sample, identifying two types of defect sites: D1 with a dark center and D2 with a bright center. Set point: -1 V, 50 pA. Scale bar: 20 nm. **d**, Large-scale STM image of the sample after 1 minute of air exposure. Set point: -1.5 V, 20 pA. Scale bar: 20 nm.

In conclusion, we have successfully demonstrated telecom-wavelength single photon emission in multi-layer InSe. By controlling strain and layer thickness, we achieved localized, bright emissions across the telecom bands. DFT and STM studies revealed that strain-induced exciton trapping and selenium vacancies are key factors in the formation of these SPEs. The direct and highly layer-dependent bandgap of multilayer 2D MMCs such as InSe makes them more suitable for layer-number engineering compared to 2D TMDCs which require monolayers for strong, direct optical transitions. We found that the layer-dependent emission is maintained and even more pronounced in the SPE forms. Our results highlight the potential of multilayer 2D MMCs as robust, tunable platforms for developing SPEs across a broad spectral range. Looking ahead, we anticipate improvements in SPE purity and fabrication yield through the growth of high-quality InSe single crystals and the deterministic implantation of deep defect states.

Method

**Sample preparation**. InSe flakes were mechanically exfoliated from a bulk crystal and subsequently transferred onto pre-patterned substrates. The layer numbers were identified using a combination of optical contrast and AFM measurements. To prepare the strain inducers, a 50 nm Au layer was deposited on a $Si/SiO_2$ substrate to block silicon PL emission. This was followed by spin coating a ~150 nm polymethyl methacrylate (PMMA) layer. The PMMA layer was then patterned into nanopillar arrays with a 5 μm pitch width using electron beam lithography. Each pillar has a height of ~150 nm and a diameter of ~200 nm. Vacuum annealing at 100°C was performed to enhance the contact between the 2D flakes and the nanopillars.

**Optical characterization.** A diagram of our optical measurement setup is presented in Supplementary Fig. S5. Micro-PL measurements of SPEs were performed using a home-built confocal microscope, with excitation provided by an 800 nm supercontinuum pulsed laser. The typical excitation power was a few µW. Samples were mounted in a continuous flow cryostat and cooled to 10 K using liquid helium. The emitted light was collected through a ×50 infrared objective lens (Olympus, 0.65 NA) and spectrally filtered before entering a 2D InGaAs array detector (NIRvana 640LN, Princeton Instruments). We used 150 and 300 gr/mm gratings to resolve the spectra. For TRPL and HBT experiments, the emission signal was spectrally filtered before being coupled into a 50:50 optical fiber beamsplitter, which evenly split the signal into two beams and directed them into two channels of a superconducting nanowire single-photon detector (Quantum Opus). PL intensity time traces, PL decay curves, and $g^{(2)}$ traces were obtained from photon detection events recorded by a PicoQuant HydraHarp 400 time-correlated single-photon-counting module.

**DFT calculations.** Plane-wave DFT calculations were carried out using the Vienna *ab initio* simulation package (VASP, version 5.4.4) with projector augmented wave (PAW) pseudopotentials for electron-ion interactions[41] and the Perdew-Burke-Ernzerhof (PBE) functional for exchange-correlation interactions.[42] Van der Waals (vdW) interactions were included using the vdW density functional method optB86b-vdW.[43] For bulk InSe, both atomic positions and lattice constants were optimized until the residual forces below 0.01 eV/Å, with a cutoff energy set at 350 eV and an 18×18×2 k-point sampling. The optimized lattice constants are a = b = 4.038 Å, and c = 25.047 Å. Single- and few-layer InSe systems were then modeled by a periodic slab geometry, where a vacuum separation of 22 Å in the out-of-plane direction was used to avoid spurious interactions with periodic neighbor cells. For the 2D slab calculations, 18×18×1 k-point samplings were used, and all atoms were relaxed until the residual forces were below 0.01 eV/Å. The in-plane lattice constants were also optimized using the setting of ISIF=4 in VASP. For 12L InSe, biaxial tensile strains ranging from 0% to 2% were introduced, and then all atoms were again optimized. Finally, we note that although the semi-local PBE functional underestimates band gaps, it is still able to well describe the band characteristics of InSe just like the hybrid functional.[44] The band gaps were corrected by using a "scissors operator" to rigidly shift the conduction bands by 0.96 eV to achieve a decent match with experimental values.

**STM and STS**. The STM/STS measurements were conducted with an Omicron VT-STM system operated at ultrahigh vacuum (UHV) conditions (P <$10^{-10}$ Torr). To reveal the native defects, single crystals of InSe were cleaved in situ at room temperature and then transferred directly to STM for imaging and spectroscopy measurements. After the first set of STM experiments, the samples were temporarily transferred to a load lock chamber and exposed to ambient conditions for one minute. Subsequent STM images were conducted at the same region of the sample examined during the first set. Standard lock-in methods were used in STS measurements with a modulation frequency of 500 Hz and modulation voltage of 50 mV (peak to peak).

**Acknowledgement**

The STM, DFT, data analysis, and manuscript writing were conducted at the Center for Nanophase Materials Sciences, which is a U.S. Department of Energy Office of Science User Facility. The sample preparation and optical measurements were performed at the Center for Integrated Nanotechnologies, an Office of Science User Facility operated for the U.S. Department of Energy (DOE), Office of Science, by Los Alamos National Laboratory. H.Z., B.L., S.J., A.-P.L, and H.H. acknowledge the support by Quantum Science Center, a National Quantum Information Science

Research Center supported by Office of Science, U.S. DOE. H.Z. was also supported in part by Wigner Distinguished Staff Fellowship at the Oak Ridge National Laboratory. L.L. acknowledges computational resources of the Compute and Data Environment for Science (CADES) at the Oak Ridge National Laboratory, which is supported by the Office of Science of the U.S. Department of Energy under Contract No. DE-AC05-00OR22725. J.C. and X.Y. acknowledge the support by Army Research Office under grant No. W911NF2410080.